\documentclass[]{interact}

\usepackage{epstopdf}
\usepackage[caption=false]{subfig}

\usepackage[numbers,sort&compress]{natbib}
\bibpunct[, ]{[}{]}{,}{n}{,}{,}
\makeatletter
\def\NAT@def@citea{\def\@citea{\NAT@separator}}
\makeatother

\theoremstyle{plain}

\theoremstyle{definition}

\theoremstyle{remark}

\begin{document}

\title{Building bridges: matching density functional theory with experiment}

\author{
\name{D.~R. Bowler\textsuperscript{a,b,c}\thanks{Email: david.bowler@ucl.ac.uk}\thanks{ORCID:0000-0001-7853-1520}\thanks{Twitter:@MillionAtomMan}}
\affil{\textsuperscript{a}London Centre for Nanotechnology, 17-19
  Gordon St, London WC1H 0AH, UK;\\
  \textsuperscript{b}Dept. of Physics \& Astronomy, UCL, Gower St,
  London WC1E 6BT, UK; \\
  \textsuperscript{c}WPI-MANA,
  National Institute for Materials Science (NIMS), 1-1 Namiki,
  Tsukuba, Ibaraki 305-0044, Japan}
}

\maketitle

\begin{abstract}
We will discuss the key concepts in density functional theory (DFT),
how it can be used to model experimental data, and consider how the
synergy between DFT and experiment can give significant insights.  The
discussion will centre on the scanning tunnelling microscope (STM) and
surface problems, tracking the author's personal interest, though the
general principles are widely applicable.
\end{abstract}


\section{Introduction}

Over the last fifty years, as modern computing has emerged and
scientific specialisms have increased, the traditional division of
experimental and theoretical physics has gained a third category,
computational physics, which has gained steadily in importance.  As a
discipline it bridges between theory, often too complex to admit
analytic solutions, and experiment, where insight from theory is
required to interpret results.  The implementation of a complex theory
as an efficient computer program which is capable of generating useful
results is challenging; understanding the limitations of the resulting
program is also key to its appropriate use.

In the realm of atomistic simulations (covering problems as diverse as
physics, chemistry, materials, earth sciences and biochemistry) the
most accurate results are provided using a quantum mechanical approach
to modelling the interactions between atoms, but this is an archetype of a
theory that cannot give analytic solutions to realistic problems.  The
most commonly used approach to finding practical, approximate solutions is known as
density functional theory (DFT), which has become ubiquitous in many
fields.  It is sufficiently mature that there are various commercial
packages that can be used to generate plausible output, with little
understanding of the reliability of the results.  Moreover, there are
a number of textbooks covering the theory and application in more
detail than is possible here, such as
Refs.~\cite{Parr:1994if,Martin:2004ja,Giustino:2014ki,Kaxiras:2003np}.

As computers have become more powerful and approaches more efficient,
the size of systems  that can be addressed (in this case characterised by the number of
atoms)has grown to the nanoscale.  At the same
time, experimental probes have become more sensitive, and can address
length scales which are on the same level as modelling.  Many
experimental probes involve some form of averaging over a relatively
large area (for instance diffraction methods), which requires a very
different approach to those methods that give atomic-scale, real-space
resolution.  The scanning tunneling microscope (STM) gives atomic
detail on metallic and semiconducting surfaces, though it involves a
convolution of atomic and electronic structure which is often
challenging to unravel.  This is where a close collaboration between
experiment and modelling is valuable: it can enable insights that the
individual disciplines cannot give.  However, it can take time, with
both sides needing to learn the capabilities and restrictions of the
other, as well as different ways of thinking and talking.

Over the course of twenty years research in this area, I have found
that the combination of DFT modelling with atomic-resolution STM gives
fascinating insight into a variety of systems and problems.  It is
not, however, a simple process to generate successful collaborations,
and requires a willingness on both sides to learn the advantages and
limitations of respective approaches.  I will describe the basic
theory behind DFT before discussing some of its limitations and
the approximations that must be made to implement it successfully.  I
will then describe two systems where the close connection between
experiment and modelling has enabled insights that go well beyond what
might be found with either approach alone, or even loosely coupled.  I
hope that the insights found here will encourage a further engagement
with the challenging but rewarding question of collaboration.

\section{Density Functional Theory}
\label{sec:dens-funct-theory}

The key equation in a quantum mechanical description of any system,
the Schr\"odinger equation, is one of the simplest to write: $\hat{H}
\vert \Psi \rangle = E \vert \Psi \rangle$; it is, however, impossible
to solve exactly for any system more complex than a hydrogen atom.
Density functional theory is one of many ways to simplify the
complexity.  Let us start without approximation, considering what makes up the
Hamiltonian, $\hat{H}$, for a system of atoms:
\begin{eqnarray}
  \label{eq:1}
  \hat{H} &=& \hat{T}_{ions} + \hat{T}_{e} + \hat{V}_{II} +
              \hat{V}_{eI} + \hat{V}_{ee}\\
  &=& -\sum_{I} \frac{\nabla^{2}_{I}}{2M_{I}} -  \sum_{i}
      \frac{\nabla^{2}_{i}}{2} + \sum_{i, j>i} \frac{1}{\mid \mathbf{r}_{i} -
                   \mathbf{r}_{j} \mid} 
      + \sum_{I, J>I} \frac{Z_{I}Z_{J}}{\mid
                   \mathbf{R}_{I} - \mathbf{R}_{J} \mid} -\sum_{I,i} \frac{Z_{I}}{\mid \mathbf{R}_{I} - \mathbf{r}_{i} \mid},\nonumber
\end{eqnarray}
where $\mathbf{r}_{i}$ and $\mathbf{R}_{I}$ indicate electron and ion
positions, respectively, $M_{I}$ is ionic mass and $Z_{I}$ is the
charge on the ion.  The terms are simply the kinetic energies of the ions and electrons
(at this stage we are treating the system with a single many-body
wavefunction, and both ions and electrons are quantum objects), and
the electrostatic interactions between the different parts of the
system.  (We work in atomic units where $q_{e} = \hbar = m_{e} =
4\pi\epsilon_{0} = 1$.)  However, as it stands, this requires a
many-body wavefunction dependent on the positions of both the ions and
electrons to be found.

In simplifying this problem, we can first note that the proton mass is
nearly 2,000 times that of the electron, while the electrostatic
forces between the particles are of similar magnitude.  As a result,
the change of the velocities of the nuclei on any timescale will be
three orders of magnitude smaller than for the electrons.  Providing
that the initial velocities are not dissimilar, it is a good
approximation to neglect the motion of the nuclei when considering
electron motion (equivalently we can say that the electrons will be in
a well-defined state---normally the ground state---when considering the
nuclei).  This is the Born-Oppenheimer
approximation\cite{Born:1927fk}, which allows us to decouple the motion of the
electrons and ions (further discussion of the approximation, its
implications and a plausible derivation can be found in Chapter V of
Ref.~\cite{Ziman:2001oq}).  An associated approximation that is almost
inevitably made is to treat the nuclei as classical: for most
simulations this is perfectly reasonable, though the mixed
classical-quantum nature of the resulting simulation raises
interesting formal (or even philosophical) questions; this
approximation is often known as the Ehrenfest
approximation\footnote{The Ehrenfest \emph{method} which is used to
  explore non-adiabatic effects\cite{Horsfield:2006ct}---that is when electrons exert an
  influence on the ions beyond their ground state potential---is a
  very different approach.}.  The forces on the ions are found from
a mean-field approximation, using the electronic charge density.

However, we are still left with the complex task of finding the
many-body wavefunction that describes the electronic ground state of
the system.  There are many approaches to this problem (which, since
we have removed the ionic degrees of freedom, now cannot
be solved exactly for a system more complex than helium); broadly they
are divided into two classes which are often described as wavefunction
methods and density methods.  The simplest wavefunction method is the
Hartree-Fock method, which represents the many-body wavefunction as a
Slater determinant of single-particle wavefunctions.  However, the
density methods are what will concern us here, specifically density
functional theory (DFT).

A model system that is often used to gain insight into more complex
electronic systems, and that forms an excellent starting point for
understanding DFT, is the uniform electron gas (which is 
characterised by the electon density).  Consider a set of
electrons contained in a box, with no potential acting on them; they
move freely, and the density of the gas can be taken to be the same
everywhere.  If we neglect the interactions between the electrons,
then the only contribution to the energy is from the kinetic energy of
the electrons.  Whatever conditions we apply at the boundaries of the
box (closed--reflecting the electrons--or periodic),
the electronic states will be characterised by wavevectors, $\mathbf{k}$,
which also give the kinetic energies of the states as $k^{2}/2$.  The
total energy is then simply the sum over the number of occupied
states, or in the limit of a very large system, the integral over
these states; in this case it will depend on the wavevector of the
highest occupied state.
It is easy to show that the number of states that are occupied, and
hence the total energy of the electronic system, depends only on the
electron density of the system. (In a three dimensional system, this dependency is
given by $E_{\mathrm{tot}} \propto n^{{5/3}}$.)

The model as described so far is for an ideal, homogeneous system, and
lacks two key ingredients necessary for modelling realistic systems: spatial
variation and electron-electron interactions.  Spatial inhomogeneity
in the system will be provided by an external potential
$V(\mathbf{r})$, typically coming from the ions.  The electron density
will become spatially inhomogeneous in response to the potential.  The
energy due to the interaction between the density and the potential is
easily found (using the classical electrostatic interaction, $E_{V} =
\int V(\mathbf{r}) n(\mathbf{r}) \mathrm{d}\mathbf{r}$).  When the
density is not homogeneous, the
kinetic energy is more problematic, and requires some form of
approximation.  As early as 1927, Thomas-Fermi
theory (well reviewed in Ref.~\cite{Spruch:1991lg}) suggested a way
forward. The key approximation in Thomas-Fermi theory is to apply the
same approach to the kinetic energy as we did to the external
potential: to calculate
the total kinetic energy as the spatial integral over the kinetic
energy \emph{density}.  If the kinetic energy \emph{per electron} for an
electron gas of density $n(\mathbf{r})$ is
$\epsilon_{\mathrm{kin}}\left[n(\mathbf{r})\right]$ then the
kinetic energy density will be
$n(\mathbf{r})\epsilon_{\mathrm{kin}}\left[n(\mathbf{r})\right]$,
and the total kinetic energy for the
inhomogeneous system can be found by integration: $E_{\mathrm{kin}} = \int
\mathrm{d}\mathbf{r} n(\mathbf{r})
\epsilon_{\mathrm{kin}}\left[n(\mathbf{r})\right]$.  From these
formulae for the energy, and a requirement that the integrated density
give the correct number of electrons, we can
derive a condition that links the density and the potential, and solve
for the inhomogeneous density that matches the external potential.

Adding electron-electron interactions generally takes a problem from
tractable to intractable; however, the key insight from Thomas-Fermi
theory is that the ground state density for the interacting system can
be found by solving the non-interacting system with an effective
potential; the interacting system can be mapped onto a
non-interacting system.  To make the effective potential we add two further terms in
the energy: the classical electrostatic potential due to the electrons; and the exchange
interaction due to the indistinguishability of the electrons and their
fermionic nature.  The former takes the standard form, and is easily
written in terms of the electron density.  The latter can be found by
analogy with the kinetic energy
in the non-interacting case: we can find
the exchange energy per electron for an electron gas of density
$n(\mathbf{r})$, written as
$\epsilon_{X}\left[n(\mathbf{r})\right]$, which is
proportional to $n(\mathbf{r})^{1/3}$.  We then approximate the total
exchange energy by integrating over the electron gas, giving
$E_{X} = \int n(\mathbf{r}) \epsilon_{X} \mathrm{d}\mathbf{r}$.

As before, we find the ground state energy by seeking the density that
minimises $E_{\mathrm{tot}}$, subject to the correct number of
electrons.  Remarkably, it can be shown that this gives the same
condition that we found for the \emph{non-interacting} case, but with
an effective potential (made up of the sum of the external, Hartree
and exchange potentials).  So the ground state charge density of an
interacting electron system is exactly the same as that of a
non-interacting electron system with the effective
potential\footnote{Note that the effective potential will depend on
  the charge density, and so solution will require an initial guess at
  the charge density, followed by a
  self-consistent iteration until the charge density and potential are
  consistent.}.

We are now in a position to describe what has become known as
Kohn-Sham density functional theory.  The first stage involves formal
proofs by Hohenberg and Kohn\cite{Hohenberg:1964qd} which extend the
ideas we have just explored in Thomas-Fermi theory, namely that the
electron density alone is sufficient to determine the ground state
properties of the system.  The proofs showed that the ground state
energy could be written as a functional of the density\footnote{More
  formally that the ground state density uniquely determines the
  external potential, and hence the many-body wavefunction and the
  ground state energy; it also implies that the ground state energy
  can be written as a functional of the density.}, and that the ground
state energy can be found by optimising the density to give the
minimum energy value of this functional.  The resulting density gives
the correct electron density distribution of the ground state.  The energy is written as:
\begin{equation}
  \label{eq:2}
  E = \int V(\mathbf{r}) n(\mathbf{r})\mathrm{d}\mathbf{r} +
  \frac{1}{2}\int\int \mathrm{d}\mathbf{r}
  \mathrm{d}\mathbf{r}^{\prime}
  \frac{n(\mathbf{r})n(\mathbf{r}^{\prime})}{\lvert \mathbf{r} -
    \mathbf{r}^{\prime}\rvert} + G\left[n(\mathbf{r})\right]
\end{equation}
where the three terms are the interaction with the external potential,
the classical electrostatic energy of the electrons, and
$G\left[n(\mathbf{r})\right]$ is a functional that we will
discuss below.  At this stage, no approximations have been made.

The second stage takes these concepts and turns them into a practical
scheme\cite{Kohn:1965hk}.  As we saw for Thomas-Fermi theory,
we will now show that within DFT the solution to the interacting electron problem with external
potential $V(\mathbf{r})$ can be given by the same density as a
non-interacting system with an effective potential,
$V_{\mathrm{eff}}(\mathbf{r})$.  However, unlike Thomas-Fermi theory,
DFT does not use a functional for the kinetic energy, as this is
generally a
poor approximation\footnote{Efforts continue to find good approximate
  kinetic energy functionals, as this would further simplify the
  problem; this approach is known as orbital-free DFT.}; instead, we solve for the wavefunctions
of the set of non-interacting electrons, using their kinetic energy,
and building the charge density from the occupied states.
We achieve this by defining the
functional $G\left[n\right]$:
\begin{equation}
  \label{eq:3}
  G\left[n(\mathbf{r})\right] =
  T_{s}\left[n(\mathbf{r})\right]  +
  E_{XC}\left[n(\mathbf{r})\right] 
\end{equation}
where $T_{s}\left[n(\mathbf{r})\right] $ is the kinetic energy of 
the \emph{non-interacting} electrons and 
$E_{xc}\left[n(\mathbf{r})\right]$ is the exchange and
correlation energy, which we will define below.  The wavefunctions
$\psi_{i}$ are found by solving the Kohn-Sham (KS) equation:
\begin{equation}
  \label{eq:4}
  \left[-\frac{1}{2}\nabla^{2} + V(\mathbf{r}) + \int
    \frac{n(\mathbf{r}^{\prime})}{\lvert \mathbf{r} -
      \mathbf{r}^{\prime} \rvert} \mathrm{d}\mathbf{r} +
    V_{XC}(\mathbf{r})\right] \psi_{i}(\mathbf{r}) = \epsilon_{i} \psi_{i}(\mathbf{r})
\end{equation}
with $V_{XC}(\mathbf{r}) =
d\left(n\epsilon_{XC}[n]\right)/\mathrm{d}n(\mathbf{r})$, and
setting:
\begin{equation}
  \label{eq:5}
  n(\mathbf{r}) = 2\sum_{i=1}^{N/2} \lvert \psi_{i}(\mathbf{r})\rvert^{2}
\end{equation}
for $N$ electrons, assuming no spin-polarisation in the system.  As
with Thomas-Fermi theory, it is important to note that this is a
self-consistent process; the generation of the Kohn-Sham wavefunctions
$\psi_{i}$ requires an input charge density.

The Hohenberg-Kohn-Sham theory gives a remarkably simple approach to the many-body problem,
with the only remaining question being: what is the functional
$E_{XC}\left[n(\mathbf{r})\right]$ ? This is a problem that cannot
be solved exactly: the form is not known analytically, and study of
how to approximate it is a hugely active field.  Recall that the KS
approach given above does not write the kinetic energy in terms of the
density, but instead calculates the kinetic energy of the non-interacting electrons,
meaning that the \emph{difference} between the interacting and
non-interacting kinetic energies in included in $E_{XC}$.  As well as
this contribution to the many-body energy, $E_{XC}$ also includes the exchange and correlation
energies, which are purely quantum mechanical in their origin.
We will discuss some of the most common functionals
developed in Sec.~\ref{sec:density-functionals}.

\section{Implementation}
\label{sec:implementation}

The transition from mathematical formalism to practical calculation
involves decisions both of implementation (in the creation of the
computer program) and instantiation (the
selection of the actual calculation performed).  These decisions will
affect the accuracy of the final result, and need to be made
carefully\cite{Brazdova:2013qt}.  In this section, we will discuss the
issues that arise in performing DFT calculations; in particular, the
various approximations that are involved should become clear.

Any simulation can only ever address a finite number of atoms; for the
simulation of an isolated molecule, this poses little problem, but for any
other simulation, some decision about the extent of the simulation must be
made.  The resulting area of space is commonly known as the simulation
cell\footnote{This is a better name than unit cell, which risks
  confusion with the fundamental units of periodic crystals.}.  A
simulation cell must be large enough to enclose the area of interest;
it must be small enough that the result can be calculated in
reasonable time.

The boundary conditions imposed on the simulation commonly take two forms: open
(where the wavefunctions match a particular value at some distance,
typically zero at infinity) and periodic (where the simulation cell is
repeated infinitely, and the wavefunction at one boundary matches its
value and derivative at the opposite boundary).  The first of these is
mainly used for molecular systems, while the second approach
enables us to use the full power of Bloch's theorem for periodic
electronic systems, and is found in condensed matter calculations of
all kinds.  Modelling a non-periodic problem, such as a defect in the
bulk of a material or a surface, with periodic boundaries requires a
reworking of the system. With a defect, it must be surrounded by
sufficient perfect bulk material to isolate it from the periodic
images of itself; a surface is typically modelled with a slab of
material surrounded by vacuum.  The choice of symmetric surfaces or a
bottom surface terminated in some inert way will depend on the details
of the system being modelled.  Choices made here will determine the
reliability as well as the accuracy of the result, and the
computational time required (better accuracy always involves more time).

When modelling the interaction between atoms that make up a material,
we might ask whether all the electrons are needed.
The understanding of bonding in chemistry suggests that only some
electrons play a significant role in bonding (valence electrons) while
others hardly change from their state in the isolated atom (core
electrons).  These core electrons are hardly involved in
the interesting or important electronic structure of molecular or
condensed matter systems\footnote{There is no formal
  definition of what forms a core electron, but these are generally
  taken to be electrons that are sufficiently strongly bound to play
  no part in bonding, and to be almost unchanged by the change in
  environment from atom to condensed phase.  Of course there are situations where
  they are important, and are involved in experimental probes such as
  X-ray absorption.}.  It seems reasonable, therefore, to develop
approaches that remove or ignore the core electrons; these will have
the advantage of reducing the number of electronic states that need to
be found.

The pseudopotential method of condensed matter physics is a standard
approach to the question of how to remove the core electrons; the
nuclear potential, along with the electrostatic potential from the
core electrons, are combined to form a new potential which is an
approximation to the full atomic potential: a pseudopotential.   This
has the added advantage that the pseudopotential is softer than the
atomic potential: it has lower curvature, and hence kinetic energy.

Pseudopotentials (sometimes PP) have been developed for over forty years, and have
progressed from rather approximate forms to extremely sophisticated
modern forms.  As we are replacing one potential with another, there
will always be an approximation made, and it is important to
understand the quality of a pseudopotential; in particular, as the
pseudopotential is generated for an isolated atom or ion, the
transferability to different environments is key.  The most common
varieties of pseudopotentials found in modern electronic structure
are: norm-conserving, or NC (where the integral of the pseudo-wavefunctions
within a specified radius gives the same number of electrons as the
all-electron calculations); ultrasoft, or US (where the norm conservation is
relaxed, giving potentially smaller basis sets and faster
calculations); and projector-augmented waves, or PAW (where the core
electrons are present but frozen in all-electron states).  All these
forms of pseudopotential can be accurate if generated and used with
care, and there have been recent efforts to test the accuracy of
different libraries of pseudopotentials that are available\cite{Lejaeghere:2016yu}.

The implementation of a quantum mechanical approach to modelling the
electronic structure of materials requires the
choice of a basis set to represent the wavefunctions; inevitably, the
basis set will be incomplete, and the balance between convergence with
respect to basis set and computational time is necessary.  Electronic
structure calculations most commonly use two broad classes of
function: delocalised functions, exemplified by plane waves
($e^{i\mathbf{k}\cdot\mathbf{r}}$, where $\mathbf{k}$ is a
wavevector); and localised functions, such as
gaussian functions or atomic orbitals (or, in the case of a
pseudopotential calculation, pseudo-atomic orbitals).

Plane waves are most widely used in condensed matter approaches: while
they are not a well-conditioned choice for condensed phase systems,
they are computationally very efficient and are well suited to
periodic systems.  In combination with PAWs or US pseudopotentials, they are
extremely efficient and reliable.  A variation of this basis set adds
extra, localised basis sets (a process known as augmentation) to give
the family of augmented plane waves; for bulk systems, the FLAPW
method (full potential linearized augmented plane wave) is one of the most accurate reference methods
available, and is used as the basis for the tests of the accuracy of
pseudopotential libraries\cite{Lejaeghere:2016yu}.

The basis size for plane waves is determined by the smallest and
largest wavevectors.  The smallest wavevector is set by the size of the simulation cell.
The largest must be determined by the user, and is almost invariably
set by reference to the equivalent kinetic energy (in atomic units,
this is given by $k^{2}/2$), frequently known as the plane wave
cutoff.  This clarifies the use of pseudopotentials: a smoother
potential means a lower plane wave cutoff.  Plane waves have the
convenient feature that the completeness 
of the basis set can be increased systematically simply by increasing
the cutoff.
 
Localised orbitals generally conform to the spherical symmetry of the atom,
and use a radial function multiplied by appropriate spherical harmonics.
The radial functions vary; the two most common forms are: gaussian
functions ($e^{-\alpha r^{2}}$), 
possibly with multiple gaussians combined into a single radial
function; and the eigenstates of the isolated atom or pseudo-atom, often
lightly confined.  The
main drawback of a localised orbital basis set is that there is no
clear systematic way to increase the size of the basis; there are
approaches (for instance, in quantum chemistry, the correlation
consistent sets\cite{Dunning:1989ys}, and the
numerical orbitals used in the AIMS code\cite{Blum:2009pi}) though
these rely on extensive development and characterisation.  Other basis
sets are used in DFT, though not as extensively, for instance: a 
discrete grid representation coupled with finite differences for
kinetic energy; and finite elements.

The computational resources required for DFT calculations depend
strongly both on the approach chosen (including the basis set and the functional) and on
the system being studied; the user should be concerned both with
computational time, and memory.  Different parts of the calculation
scale differently with system size (generally characterised by the
number of atoms, $N$) but there are limiting factors: ultimately the
computer time will scale with the cube of the system size (or
$\mathcal{O}(N^{3})$ scaling) with the memory scaling with its
square.  This behaviour tends to limit calculations to systems of tens
or hundreds of atoms, with very few passing a thousand atoms.

The cost will also depend on the accuracy required: in periodic
systems, the sampling of
the Brillouin zone (which reduces with increasing system size); any
numerical grid required for integration in real space; the size of the
basis set chosen; and which electrons to include (as valence electrons) in the
calculation.  Particularly for grid and plane wave basis sets, the
cost can depend strongly on the element: first row elements and first row
transition metals often require a larger basis (the ion core is less
screened than for other elements, giving larger kinetic energies).

\subsection{Larger scale calculations}
\label{sec:larg-scale-calc}

Where does the scaling of computer effort and memory with number of
atoms originate ? The DFT eigenstates extend over all space
(in practice, over the whole simulation cell).  This spatial extension is
responsible for the quadratic scaling of memory with system size\footnote{The
number of electronic states depends on the number of atoms, and the
amount of information in each state depends the volume of the system,
which in turn depends on the number of atoms.}, and
the cubic scaling of computational effort\footnote{For most methods,
  the cubic arises through the
  need to orthogonalise the eigenstates, requiring an integral which
scales with system size alongside the $N^{2}$ pairs of eigenstates;
local orbital methods that directly diagonalise the Hamiltonian incur
the cubic cost in the diagonalisation step}.  
Large-scale calculations, and long time-scale calculations, can be
performed more quickly using parallelisation: multiple processors are
each assigned responsibility for different parts of the calculation.
This area is typically known as high-performance computing (HPC), with
facilities ranging from powerful workstations (a single machine with
several tens of computational cores) to national centres with
tens or hundreds of thousands of cores.

There are limitations on the efficiency of parallelisation:
certain operations scale poorly with numbers of processors, while
other operations require significant communication between
processors.  The fast Fourier transform (FFT) is used extensively in
codes with plane wave basis sets; while exceptionally efficient
implementations exist, at large system sizes it often presents a
significant bottleneck.  The largest calculation to date using
standard DFT was demonstated on 100,000
atoms\cite{Hasegawa:2011tu,Hasegawa:2014vk}, though just one
self-consistency cycle was performed, and the calculation used used
400,000 processor cores; the same real-space code has demonstrated
practical calculations on up to 10,000 atoms, but very few DFT
calculations go beyond 1,000 atoms.  Use of local basis sets can
improve performance, largely because operations can be made local in
space so that matrices become sparse.
  
For insulating and
semiconducting systems,  however, it can be shown that the electronic
structure is localised in real-space\cite{He:2001cq} (this has been
referred to as the ``near-sightedness'' principle\cite{Kohn:1996lp}).
Since the relevant information for any given point or atom can
therefore be confined to a local volume, it should be possible to
formulate DFT calculations in a linear scaling, or $\mathcal{O}(N)$,
fashion\footnote{It is worth noting that, as seen above, the original principle of
  DFT, that the ground state energy is a functional of the density,
  also implies the same scaling; it is the practical calculation of
  the kinetic energy that prevents the use of the charge density in
  general; the orbital-free DFT methods seek to address this, building
  from Thomas-Fermi theory.}.  There
are indeed a significant number of linear scaling methods, and several
codes that implement these\cite{Bowler:2012zt}.  To date, the largest
calculations have been demonstrated on over a million
atoms\cite{Arita:2014qr,Bowler:2010uq}, though the stable, accurate
implementation of these techniques is a challenge.

\subsection{Density functionals}
\label{sec:density-functionals}

We have so far steadfastly avoided discussions of what density
functionals are actually used; here we will give a \emph{brief}
overview of the classes of functional, and their most commonly used
implementations.  There are many reviews that can be read in this
area, and we can only suggest a
selection\cite{Becke:2014bz,Cohen:2012cf,Yu:2016sz,Mardirossian:2017hl}.

The simplest approximation, suggested in the original derivation
of DFT\cite{Kohn:1965hk}, is the local density approximation (LDA),
where the exchange and correlation energy is written as:
\begin{equation}
  \label{eq:7}
  E_{XC}\left[n\right] = \int \mathrm{d}\mathbf{r} n(\mathbf{r}) \epsilon_{XC}[n(\mathbf{r})]
\end{equation}
where $\epsilon_{XC}[n(\mathbf{r})]$ is the exchange and correlation
energy \emph{per electron} for a uniform electron gas of density
$n(\mathbf{r})$.  The exchange functional can be written analytically,
while the correlation functional is fitted to exact results and
calculations for the uniform electron gas from accurate methods such
as quantum Monte Carlo\cite{Ceperley:1980ch}; a number of
parameterisations exist, such as Ref.~\cite{Perdew:1992xp}.

The resulting approximation is quite simple, and surprisingly
effective; it is the only approximation possible that depends solely on
the density (and thus is a local approximation).  However, there are various failures and restrictions
associated with this simplicity, discussed in the next section, which
have provoked the development of functionals with more complex
dependency on the density (e.g. gradients, giving semi-local
functionals, etc.).  The ultimate aim,
the exact functional, can be proven to exist\cite{Hohenberg:1964qd} but
is not known exactly (moreover, it is not known as the end point of a series of
approximations converging systematically\cite{Perdew:2005yh}; while it
is possible to calculate the 
exact answer for specific small systems, this is more expensive than
solving the Schr\"odinger equation exactly\cite{Stoudenmire:2012jl}).

One of the key members of the field, John Perdew, coined the idea of a
Jacob's ladder of functionals\cite{Perdew:2001of}, with successive
rungs each more accurate\footnote{Based on the biblical story of Jacob in Genesis
  chapter 28, where he sees a ladder or staircase reaching to heaven;
  given the huge effort expended and many thousands of functionals
  developed, another eminent scientist in the field, Mike Gillan, suggests that a
more apt biblical metaphor might be wrestling Jacob, from four chapters later, where Jacob
wrestles all night with God.}.  The increase in accuracy proceeds in
stages, each of which typically increases the computational cost:
\begin{enumerate}
\item The local density approximation (LDA), depending only on the
  value of the density at each point in space, $n(\mathbf{r})$
\item Generalised gradient approximations (GGA), where the functional
  also depends on the magnitude of the gradient of the density,
  $\left\vert \nabla n(\mathbf{r})\right\vert$
\item Meta-GGA functionals include the kinetic energy density:
  \begin{equation}
    \label{eq:8}
    \tau_{\sigma}(\mathbf{r}) = \frac{1}{2}\sum_{i}\left\vert\nabla\psi_{i}(\mathbf{r}\right\vert^{2}
  \end{equation}
  where the sum is over occupied Kohn-Sham eigenstates, $\psi_{i}$, only
\item Hybrid functionals add some fraction of the exact exchange
  energy written in terms of the Kohn-Sham orbitals.  There are some
  variants of hybrid functionals that introduce range-separation,
  where the exchange energy is calculated partly from the exact
  exchange and partly from GGA exchange (typically this is for solids,
  where the computationally expensive exact exchange is only used for
  short-ranged interactions)
\item The final set of functionals introduce a dependency on both
  unoccupied and occupied orbitals, using approaches such as the
  random phase approximation (RPA), but these are extremely expensive
\end{enumerate}
There are other additions (notably the van der Waals density
functionals discussed below) but these are the key rungs on the
notional ladder of accuracy.  The functional chosen depends on the
type of calculation being performed.  Most solid state calculations
use a GGA functional (in part because most use plane waves as a basis
set, and this makes hybrid calculations very expensive) though
Meta-GGA is becoming more common.  Many quantum chemistry calculations
use a hybrid functional.  Some of the most commonly used functionals include:
the GGA functional PBE\cite{Perdew:1996su} and its variants; the
hybrid functionals B3LYP\cite{Becke:1993zv} and
PBE0\cite{Perdew:1996ie}; the screened hybrid HSE\cite{Heyd:2003pz};
and the meta-GGA functionals TPSS\cite{Tao:2003kl} and
SCAN\cite{Sun:2015wb}.  The selection and reliability of a functional
depends both on the system being studied and, to some extent, the
philosophy of the user.  Approaches to functional creation include
on the one hand ensuring that the functional conforms to known
constraints and behaviours, and on the other hand fitting parameters
in the functional to databases of accurate existing results.  These
topics are discussed further
elsewhere\cite{Burke:2012pw,Pribram-Jones:2015ci,Yu:2016sz}. 

\section{Capabilities and Restrictions}
\label{sec:capab-restr}

We will briefly discuss some of the successes and limitations of DFT
in general, and specific functionals in this section.  There are many
reviews of this
area\cite{Becke:2014bz,Burke:2012pw,Hasnip:2014yb,Jones:2015ma,Cohen:2012cf,Pribram-Jones:2015ci}
to which the interested reader is directed.

Geometries are almost always better with GGA than LDA, which
overbinds (in the solid state, lattice constants are often around 1\%
too small).  The PBE functional\cite{Perdew:1996su} which is commonly used for GGA
calculations overestimates lattice constants by about the same amount;
the PBEsol functional\cite{Perdew:2008im} is a recalibration of the
PBE functional for solids (losing some accuracy for atomization and
total energies).

While electronic structure is often well described, the most
well-known issue with LDA and GGA is that band gaps are seriously
underestimated.  The major contribution to this issue is the
self-interaction: the expression for the electrostatic energy of the
electrons includes the interaction of each electron with itself.
Attempts have been made to create approaches within DFT that correct
for this\cite{Perdew:1981vh,Goedecker:1997fd} but are often hard to
converge and do not improve accuracy significantly.  Hybrid
calculations, where some exact exchange is included, correct this
issue while also improving thermochemistry, at significantly higher
computational cost.

The area where much recent work has been directed is that of
non-bonding interactions, specifically van der Waals forces.
Significant progress has been made with semi-empirical methods that
add extra forcefield terms to add
dispersion\cite{Grimme:2010ss,Tkatchenko:2009gx}.  However,
significant advances have been made in developing a fully non-local
DFT dispersion functional\cite{Dion:2004vp}, which has 
been extensively tested and further
developed\cite{Klimes:2012sa,Berland:2015pf}.  There now exist
excellent DFT approaches to dispersion forces.

The question of excited states is one we have not yet addressed.  The
DFT eigenvalues do not have any formal link to measured energy levels
(beyond the frontier orbitals) but are often used as an ansatz for
them.  DFT is a ground state theory: the original theorems only apply
to the ground state density, and hence the occupied levels.  The most
commonly used approach to calculated excited state energies is
time-dependent DFT (TDDFT)\cite{Marques:2004lh,Marques:2006cn}, while
many-body perturbative methods\cite{Onida:2002ek}\cite{Reining:2018id}
often build on DFT eigenvalues as their starting point.  With these
methods a wide variety of spectroscopies can be modelled and probed.

\section{Examples of collaboration}
\label{sec:exampl-coll}

While DFT is a powerful method for understanding the properties of
molecules and materials, it is much more powerful when work is
carried out in close collaboration with experiment.  It is relatively
easy as a DFT practitioner to create a new structure, phase or
material, and to publish the result.  It is more informative to select
the system or structure on the basis of experimental evidence,
narrowing the space of problems, and to provide insight into
experimental results that go beyond the resolution or capabilities of
experiment.

It is far better, though often challenging, to work in collaboration
with experimental groups, with frequent feedback and discussion,
giving rise to combined results that exceed what would have been
possible for either discipline in isolation.  The challenge in this
mode of working comes from the investment required to learn about the
strengths and limitations of the different techniques, and to
understand how the data and insight from one can illuminate and open
up the other.

This section concentrates almost exclusively on STM, as this is the
technique with which I have mainly engaged.  However, of course, there
are many experimental techniques that can be addressed with DFT,
including: NMR; TEM, especially EELS; photo-emission spectroscopy;
inelastic neutron and x-ray scattering; and both IR and Raman
spectroscopy, to name some of the most common.

\subsection{Bismuth on Si(001)}
\label{sec:bism-si001-sect}

Nanowires have been a topic of intense research recently\cite{Owen:2006mi}, and
self-assembled nanowires are very attractive from a manufacturing
point of view.  The structure of the nanowires (or nanolines) that
self-assemble on the Si(001) surface following adsorption and
annealing of bismuth required considerable time, and
a concerted collaboration between experiment and theory, to unravel.

\begin{figure}
  \centering
  \includegraphics[width=0.5\textwidth]{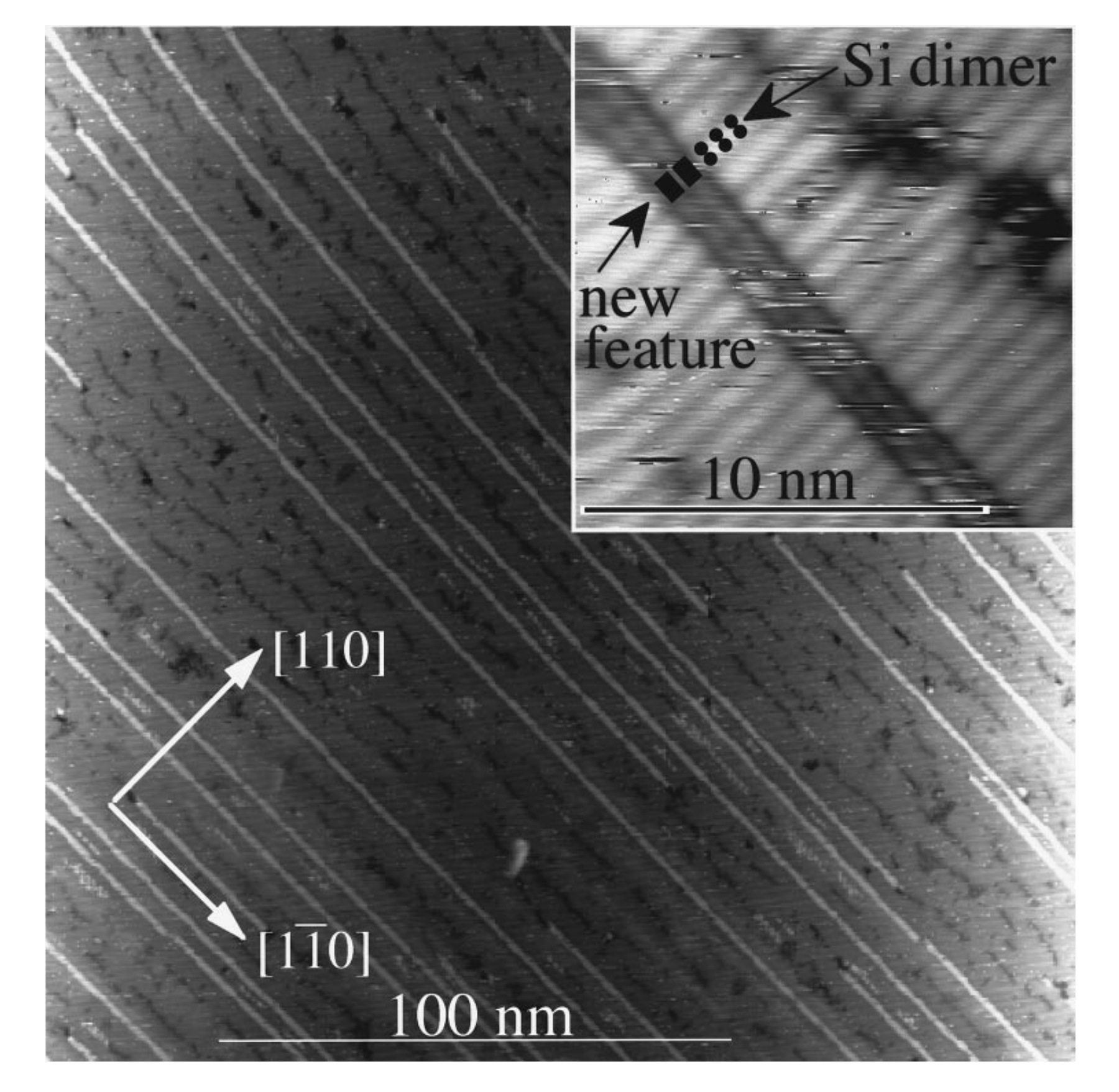}  
  \caption{STM image of Bi nanolines in Si(001), taken at +2V.  The
    inset shows a detailed scan of the nanolines, with postulated
    positions of silicon and bismuth dimers (subsequently shown to be
    in different locations). Reprinted figure with permission from
    Ref.~\cite{Miki:1999ta}. Copyright 1999 by the American Physical Society.}
  \label{fig:BiNanolines}
\end{figure}

Bismuth is used in semiconductor growth as a surfactant: a material
that passivates the surface, leading to flatter, smoother
growth\cite{Sakamoto:1993qp}.
However, it was found experimentally\footnote{The actual discovery was
  a complete accident: a sample that had been left to anneal
  overnight, with the intention of removing all bismuth, was the
  source of the first observation of the bismuth nanolines} that it
has another property on silicon surfaces: it can form atomically
perfect nanolines, 1\,nm wide and up to microns
long\cite{Naitoh:1999aq,Naitoh:1997mb,Miki:1999ta,Miki:1999ud}, as shown in
Figure~\ref{fig:BiNanolines}.  (We refer to these structures as
nanolines rather than nanowires as their gap is larger than that of
the surrounding surface.)  The initial STM images were taken at
elevated temperature, and gave little detail: the basic appearance of a
nanoline with two parallel tracks was clear, but the actual width and registry with the substrate
was not visible.  The initial structures proposed matched these basic
facts, but could not explain the key feature of these lines: their
extraordinary straightness and perfection.  A kink has never been
observed in any bismuth nanoline, while defects are very unusual.

The Si(001) surface consists of rows of dimers (pairs of Si atoms
bonded together in a surface reconstruction).  The model initially
proposed for the Bi nanolines\cite{Miki:1999ta} used
the available evidence, and assumed that two bismuth
dimers would replace three silicon dimers, with a vacancy between
them.  This early model matched the basic observations , and was
stable, but had one main problem: there was no mechanism to explain
the straightness of the lines.  At this stage, the modelling
contribution was limited by the available experimental data---the
energetics could be calculated, but without more detail on the
structure little could be done.

These original experiments were performed at elevated temperatures to
keep the surface clean\footnote{The deposition process for the bismuth
produced contaminants, and the silicon surface is sufficiently
reactive, that cooling to room temperature resulted in a very dirty
surface which was not appropriate for high resolution imaging.},
reducing the resolution of the imaging.  Passivating the surrounding
surface with hydrogen\cite{Naitoh:2000zt,Owen:2002cq} revealed that
the bismuth lines in fact replaced four silicon dimers, not three, and
were situated between the underlying surface dimers.  The first model
proposed to occupy the space of four dimers\cite{Naitoh:2000zt}
suggested two vacancies bracketing a central pair of bismuth dimers,
which failed to match the experimentally observed spacing and
location of the parallel tracks.

\begin{figure}
  \centering
  \includegraphics[width=0.5\textwidth]{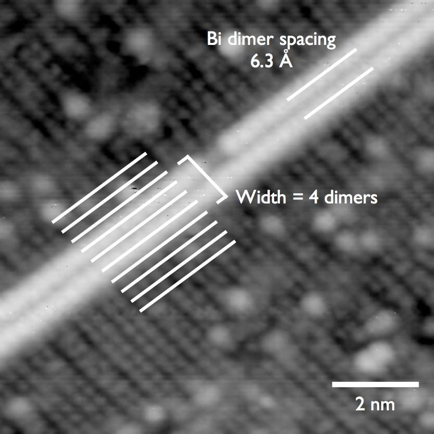}  
  \caption{Bi nanolines with passivated Si(001) background, imaged in
    STM at -2.5V and
    room temperature. Substrate dimers are shown with a set of white
    lines, which enables registry of Bi nanolines with the substrate
    to be seen eaily.  Reprinted from Ref.~\cite{Owen:2006mi} by
    permission from Springer, copyright 2006.}
  \label{fig:BiNanolinesRT}
\end{figure}

The situation at this point neatly encapsulates one of the major
problems for surface science, and condensed matter more generally: there is
limited experimental information on a surface reconstruction or
crystal structure, and a wide parameter space that could be searched
with modelling.  In recent years, approaches have been developed to
search a wide variety of structures\cite{Woodley:2008zg} (these
methods include high throughput techniques such as the materials
genome\cite{Pablo:2014wk} and random structure
searching\cite{Pickard:2011to}, and evolving approaches such as
genetic algorithms\cite{Woodley:2008zg}) but there is little
success with or application to surfaces.

We proceeded to use a combination of intuition and modelling
constrained by the known properties of the system.  The information
available was as follows: 
\begin{itemize}
\item The lines only formed near the Bi desorption temperature, after
  annealing
\item The lines persisted after desorption of the remaining surface Bi
\item No kink was observed in the lines, and almost no defects
\item The silicon around the lines was defect-free: the lines appeared
  to exclude defects
\item The lines replace four silicon surface dimers
\item The spacing between features in the line was less than the
  spacing between dimers in the surface
\end{itemize}

Based on these observations, we sought structures that would be more
stable than the simple $\left(2\times n\right)$ reconstruction of Bi
dimers adsorbed on the surface.  We used a linear-scaling tight
binding code to explore the stability of possible
structures\cite{Bowler:2002ux}; the results were approximate but very
fast, and enabled large scale calculations (1,000+ atoms) on very
modest computational resources (a laptop in the lab), as well as
making the rapid testing of many candidate structures almost trivial.  Promising
candidate structures were refined using DFT; these calculations showed
that the tight-binding results were, with one unimportant exception, correct in
the ordering of stability of structures.

\begin{figure}
  \centering
  \includegraphics[width=0.5\textwidth]{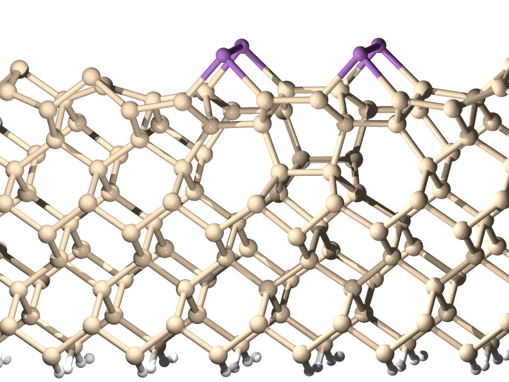}  
  \caption{The Haiku model for the Bi nanolines\cite{Owen:2002rq}, seen in perspective.
 This slab model has ten layers of silicon, with the base fixed and
 terminated in hydrogen to mimic semi-infinite bulk.  The normal
 tetrahedral bonding of silicon is clear in the lower layers.  The
 significant reconstruction below the bismuth dimers, shown in purple,
 is formed of five-membered and seven-membered rings.  The Bi nanoline
 replaces four silicon dimers in the surface.}
  \label{fig:HaikuModel}
\end{figure}

The structure that was identified as the Bi nanoline\cite{Owen:2002rq}
is shown in Fig.~\ref{fig:HaikuModel}.  It drew inspiration from STM
observations of step edges on As-covered Ge(001)
surfaces\cite{Zhang:2001ff}.  The model is named Haiku\footnote{After
  the Japanese verse form with three lines consisting of five, seven
  and five syllables, respectively.} and shows considerable
sub-surface reconstruction (with five-membered and seven-membered
rings prominent---reminiscent of defects in carbon nanotubes).  It
fits all the criteria identified for the nanolines.

\begin{figure}[h]
  \centering
  \includegraphics[width=0.5\textwidth]{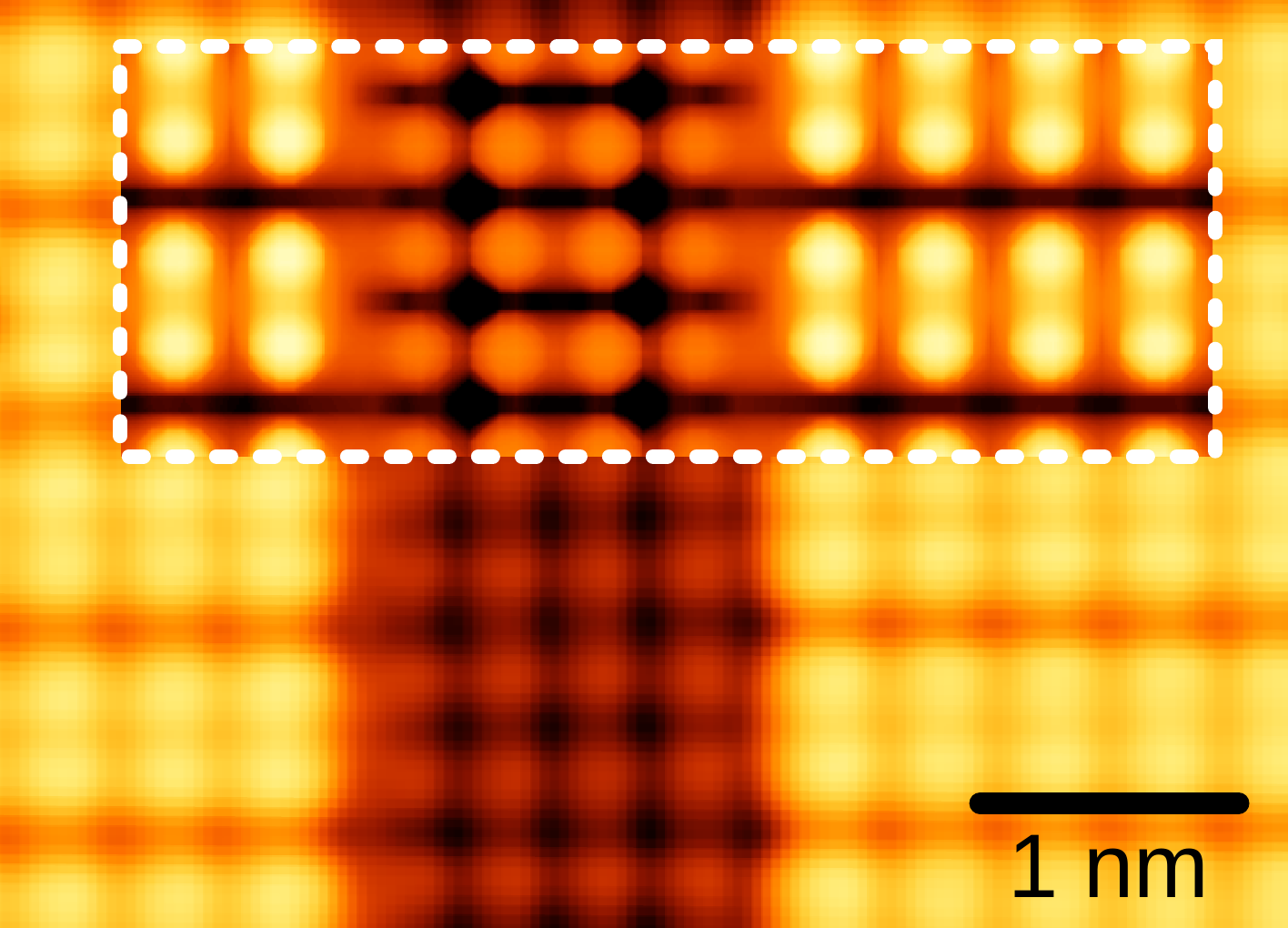}
  \caption{Main image: STM image at -2.5V of Si(001) surface after Bi nanoline
    growth and dosing with hydrogen.  The bare haiku trench runs
    vertically through the figure.  Inset (within dashed rectangle):
    Simulated STM image from DFT.  Reprinted figure with permission
    from Ref.~\cite{Bianco:2011vn}. Copyright 2011 by the American Physical Society.}
  \label{fig:BareHaiku}
\end{figure}

The ultimate confirmation of the correctness of the structure came
when it was discovered that a high flux of atomic hydrogen could
remove the Bi\cite{Owen:2010fk}.  An STM image of the resulting
structure is shown in Fig.~\ref{fig:BareHaiku}; there is a clear
difference between the surface and the area left vacant by Bi.  Inset into the image is a
simulated STM image from DFT of the Haiku structure with Bi replaced by
hydrogen atoms; the agreement is remarkable.

Further exploration of the Bi lines has found other detailed features
that required more careful collaboration between experiment and theory: a
peculiar feature observed in STM with sub-\AA ngstr\"om dimensions
was shown to be a Si dimer
substituting for a Bi dimer in the line, rapidly flipping between
different orientations\cite{Kirkham:2017xd}; and subtle changes in the
appearance of the nanoline with STM voltage was shown to result from
electronic coupling between the background silicon and the Bi
dimers\cite{Longobardi:2017kl}.  In all cases the information
available from either experiment or theory was incomplete, and the
final interpretation was only possible with careful dialogue and
exploration between the two. 

\subsection{TiSe$_{2}$}
\label{sec:tise2}

\begin{figure}
  \centering
  \includegraphics[width=0.5\textwidth]{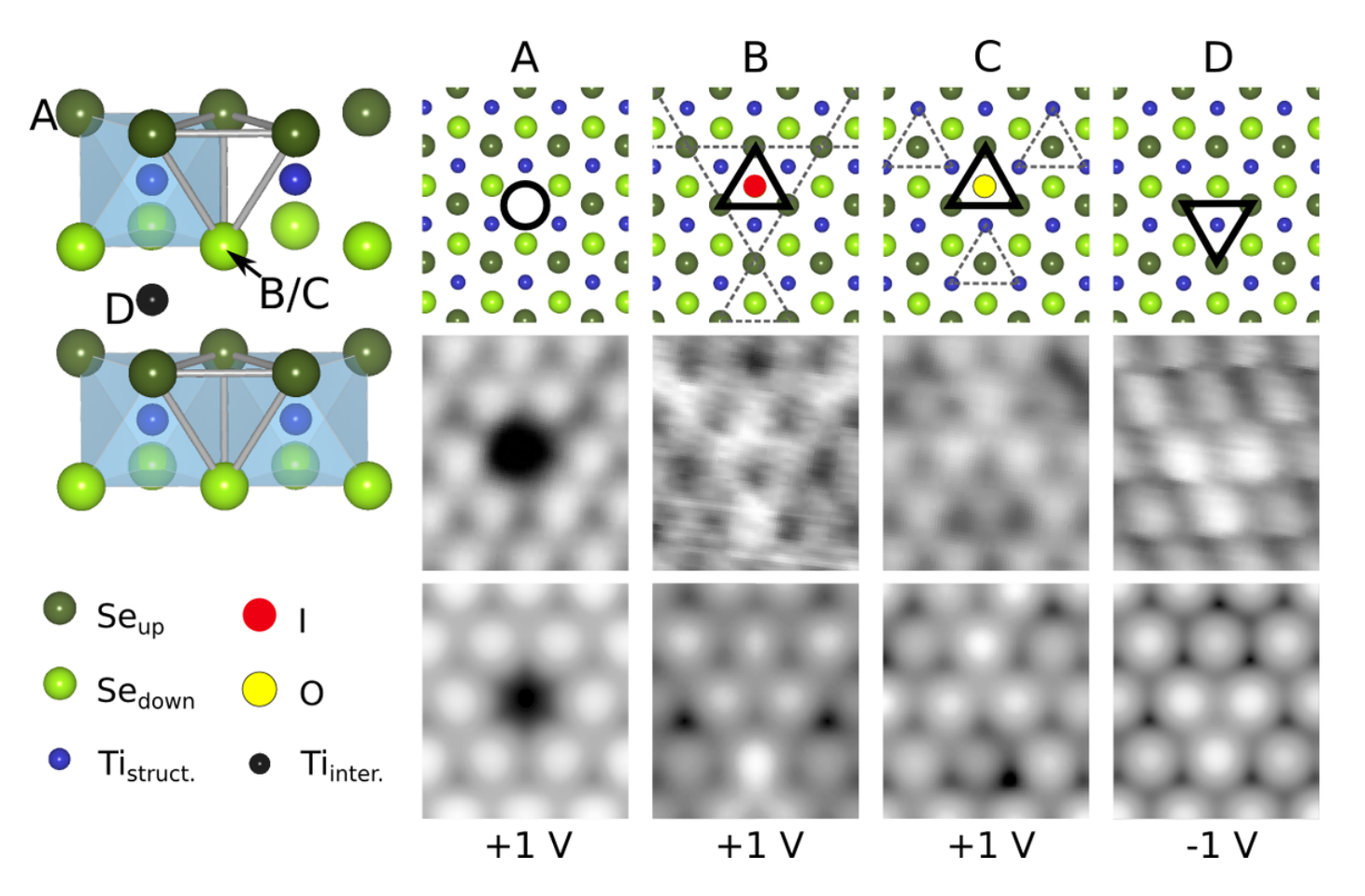}  
  \caption{Top left: schematic illustration of the structure of
    TiSe$_{2}$ and the location of the four defects.  Main: schematic,
    STM and DFT images of four defects in TiSe$_{2}$, without charge
    density wave.  Reprinted figure with permission from
    Ref.~\cite{Hildebrand:2014nj}. Copyright 2014 by the American Physical Society.}
  \label{fig:TiSe2NoCDWDef}
\end{figure}

TiSe$_{2}$ is a quasi-two-dimensional material, one of the large
family of transition metal dichalcogenides, consisting of a
sandwich of a hexagonal layer of Ti between two hexagonal layers
of Se, with successive sandwiches loosely bound
together by van der Waals forces; this is schematically illustrated in
the left-hand side of Fig.~\ref{fig:TiSe2NoCDWDef}.  The material develops a charge-density
wave (CDW) below $\sim$200K, which is commensurate with the lattice
(somewhat unusually for this class of material).  It is not clear whether it is a semiconductor
with a very small indirect gap or a semimetal\cite{Monney:2015mb,Rasch:2008nh}.  It can be made
superconducting by application of pressure\cite{Kusmartseva:2009rr} or
doping with Cu\cite{Morosan:2006pe}.  These
properties make it a very interesting system to study, particularly as
it gives considerable opportunity to gain insight into the relationship
between the formation and development of the CDW and
superconductivity.  Using STM and DFT modelling 
together gives the opportunity to study the correlation between the
CDW and the material's electronic structure in real space. 

As with all materials, there are defects in the TiSe$_{2}$ samples
that are grown.  Understanding the real-space location of these native defects in TiSe$_{2}$ in both
the non-CDW and CDW forms will give important points of reference when
studying more complex features in the system.  Similarly,
understanding the defects without the CDW is an important precursor to
understanding the effect of the CDW on the defects, and vice versa.
The information available from STM alone is not enough to identify
defects; using the appearance of the defects and how they change with
imaging bias, alongside DFT modelling of likely defect sources, is the
only way to characterise the defects.

The right hand side of Fig.~\ref{fig:TiSe2NoCDWDef} (columns A-D) shows
the four native defects identified from TiSe$_{2}$ samples grown with
iodine-vapour transport.  The middle row shows atomic resolution STM
images of the four defects identified in the samples; most (A-C) were
clearest in empty states, or positive sample bias, while one (D) was
only seen in filled states, or negative samle bias.  From these images,
and the likely contaminants or lattice defects, a list of DFT
simulations was created.  For each of these, after structural
relaxation\footnote{This simple phrase often masks a long and painful
  process, particularly in systems with very shallow energy surfaces
  in one direction.}, STM images were simulated at a variety of
voltages.  The bottom row in Fig.~\ref{fig:TiSe2NoCDWDef} shows the
best match for each STM image, while the top row shows a schematic of
the proposed defect.  Iodine (B) is present because of the growth process,
while oxygen (C) is almost unavoidable; the top layer vacancy (A) is likely to
come from desorption of Se or I/O substitutional defects following
cleaving of the crystal.  Ti intercalates (inserts into the van der
Waals gaps between layers) during the growth process.  The temperature
of the growth process controls to some extent 
the excess levels of Ti incorporated in the crystal, and the defect D
is only seen in crystals with Ti self-doping, which coupled with good
STM agreement, suggests that this defect is a Ti intercalate.  These atomically
precise identifications allow future studies to identify the registry
of STM images and the underlying substrate.

\begin{figure}[h]
  \centering
  \includegraphics[width=0.5\textwidth]{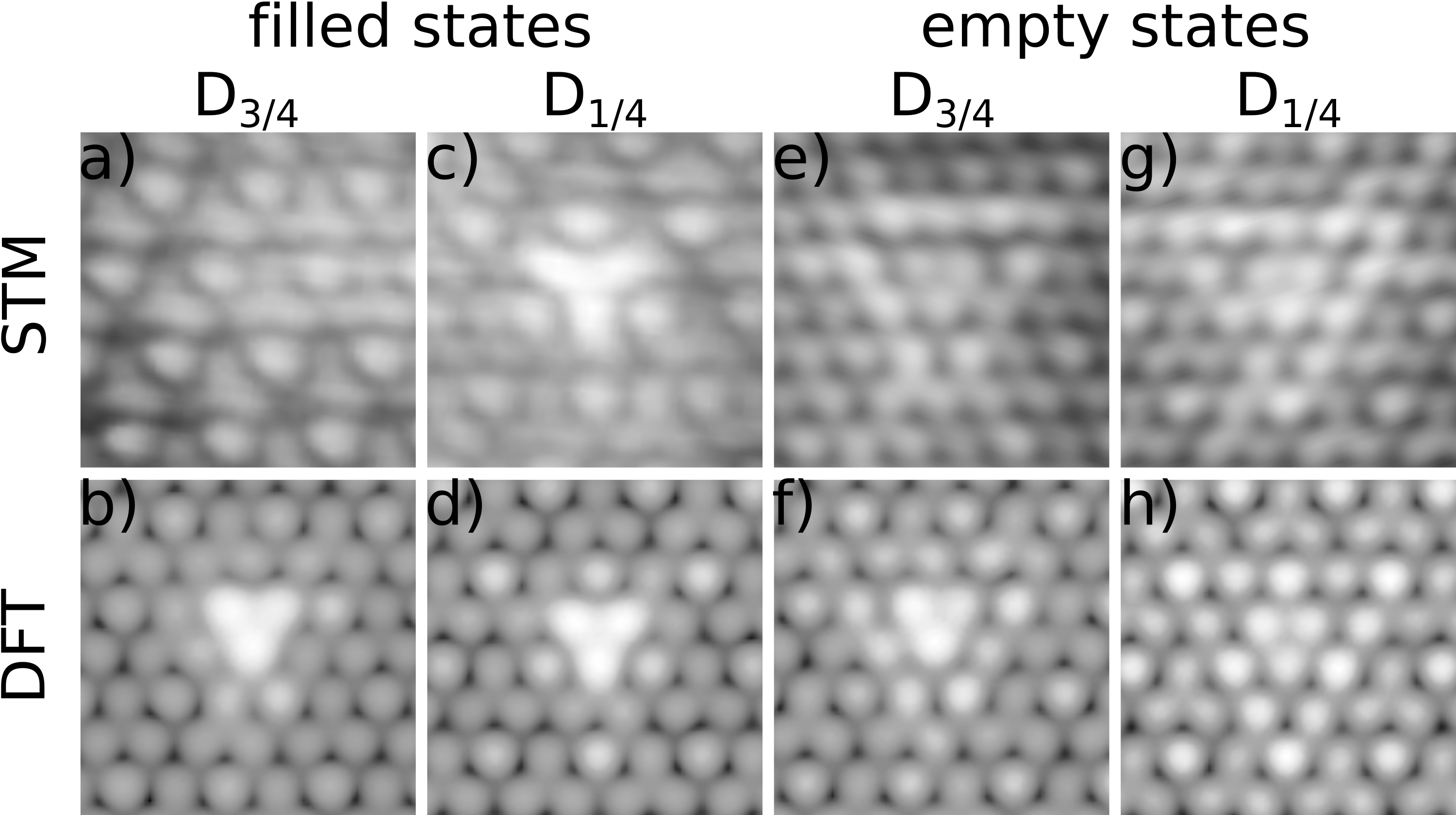}
  \caption{STM (top) and DFT simulated images (bottom) of Ti
    intercalate defects (type D in Fig.~\protect\ref{fig:TiSe2NoCDWDef}) in the presence of the CDW.
    Filled states bias was -150mV, empty states +150mV.  Reprinted figure with permission from
    Ref.~\cite{Novello:2015mc}. Copyright 2015 by the American Physical Society.}
  \label{fig:TiInterCDW}
\end{figure}

Cooling below $\sim$202K induces the formation of the charge density
wave in TiSe$_{2}$, which is only visible in STM at biases below
$\sim$0.2V.  It takes the form of a $(2\times 2)$ periodic distortion,
commensurate with the underlying lattice, with four atoms in the
surface unit cell.  The unit cell divides into one bright atom and
three darker atoms, giving two sites for the defects, labelled 1/4 and
3/4, depending on their relation to the surface atoms.  The first
three defects identified in Fig.~\ref{fig:TiSe2NoCDWDef} above, A--C,
are only slightly perturbed by 
the presence of the CDW, and can be well described in DFT without the
CDW distortion\cite{Novello:2015mc}.  The Ti intercalate, shown in
Fig.~\ref{fig:TiInterCDW}, interacts more strongly with the CDW in
terms of its appearance.  However, despite the doping nature of this
defect, it does not disrupt the CDW locally---an important observation
in the light of resistivity measurements that show the macroscopic phase
transition signature disappearing with Ti
self-doping\cite{Di-Salvo:1976cb}.  Overall, these measurements and
modelling enabled us to rule out the formation of an incommensurate
CDW, and any local change of crystal structure (for instance change
from 1T to 2H polytype), and to note that there are different
electronic signatures from the defects on inequivalent lattice sites.

\begin{figure}[h]
  \centering
  \includegraphics[width=0.48\textwidth]{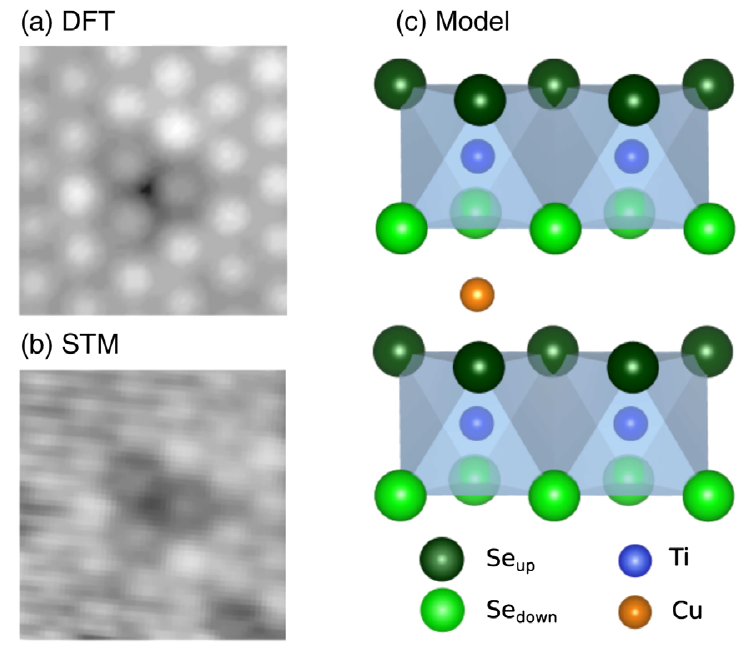}
  \includegraphics[width=0.48\textwidth]{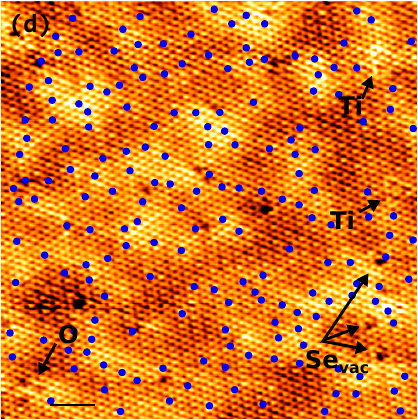}
  \caption{(a) DFT simulation and (b) STM image of an intercalated Cu
    atom (1.72$\times$1.72 nm$^{2}$, V$_{bias}$=-1.2 V, I = 30 pA). (c)
    Model of the 1T-Cu$_{x}$TiSe$_{2}$ showing the Cu atom position in the vdW
    gap. (d) Low bias image of surface showing the CDW modulation,
    with Cu intercalates indicated with blue dots (identified by
    imaging at different bias).  Reprinted figure with permission from
    Ref.~\cite{Novello:2017wq}. Copyright 2017 by the American
    Physical Society.}
  \label{fig:TiSe2Cu}
\end{figure}

The ability to interrogate, in real-space, the effect of different
defects and dopants on the CDW is extremely valuable, and gives new
insights into the stability and formation of the CDW, as well as the
relation to superconductivity.  Addition of Cu, beyond a fractional
doping of 0.04, induces superconductivity in 
1T-TiSe$_{2}$, with a maximum $T_{C}$ of 4.1K when $x\simeq
0.08$\cite{Morosan:2006pe}, with transport measurements showing that
the CDW is suppressed as the Cu fraction is increased.  However, the
location of the Cu atoms and their electronic signature were not
known.

Using DFT, we were able to predict a likely bias voltage and
appearance for the Cu intercalates, shown in Fig.~\ref{fig:TiSe2Cu}(a)
with a structural model in (c).  Subsequent STM measurements found this exact
appearance, as seen in Fig.~\ref{fig:TiSe2Cu}(b)\cite{Novello:2017wq}.
We have already established that the CDW maxima are located on Se
atoms using native defects; we can now image both the Cu atom location
(at relatively high bias, $\sim -1.2V$) and the local effect on the CDW
(at low bias), as shown in Fig.~\ref{fig:TiSe2Cu}(d).  In particular, the
CDW breaks up into nanometre-scale 
domains with short-range order, and other areas where the CDW
amplitude is reduced.  STM shows that areas with high Cu concentration
correlate with the reduction of the CDW amplitude--an insight that
would impossible without detailed knowledge of the location of the Cu
atoms, which in turn was found only by close collaboration between
experiment and modelling.

\section{Conclusions}
\label{sec:conclusions}

Density functional theory has become ubiquituous in many fields, but
is a sufficiently complex technique that some understanding of its
basic implementation in modelling packages is vital to successful
use.  Moreover, modelling on its own is futile, and requires
at the very least comparison to experiment, while collaboration is far
more rewarding.

I have given a basic outline of the theory behind DFT, as well as
discussing some of the issues involved in its implementation.  This
should highlight the dangers of using it as a black box: many pitfalls
lie in wait for the unwary user!

By focussing on two key areas of where I have had extensive
collaboration with experimental groups, I have highlighted the gains
that can come from spending significant time to learn the capabilities
and restrictions of different methods.  Understanding what can and
cannot be done both with an experimental approach and with modelling
is highly instructive.  I hope to inspire further deep collaborations.

\section*{Acknowledgement(s)}

I have worked with many skilled experimentalists and modellers over
the years, and do not have space to list them all.  I am particularly
pleased to acknowledge (in no particular order) James Owen,
Kazushi Miki, Christoph Renner, Anna-Maria Novello, Baptiste
Hildebrand, Chris Goringe, Chris Kirkham, Tsuyoshi Miyazaki and Mike
Gillan for fruitful and illuminating collaborations and discussions
over many years.

\section*{Abbreviations}

A list of abbreviations commonly used in the field, in alphabetical
order.

\begin{tabular}[h]{ll}
  BLYP & Becke (exchange) and Lee, Yang \& Parr (correlation) GGA
         functional\\
  DFT & Density functional theory\\
  FLAPW & Full-potential linearized augmented PW\\
  GGA & Generalised gradient approximation\\
  KS   & Kohn-Sham\\
  LDA & Local density approximation\\
  LMTO  & Linear muffin-tin orbital \\
  PAW  & Projector-augmented waves\\
  PBE & Perdew, Burke \& Ernzerhof GGA functional\\
  PW  & Plane wave \\
  TDDFT & Time-dependent DFT\\
  vdW  & van der Walls (also called dispersion)\\
  XC  & Exchange and correlation\\
\end{tabular}

\bibliographystyle{tfnlm}
\bibliography{TheoryMeetsExperiment}

\end{document}